# Can we improve the environmental benefits of biobased PET production through local biomass value chains? – A life cycle assessment perspective

**Carlos A. García-Velásquez[a], Yvonne van der Meer[a1]**

[a] *Aachen-Maastricht Institute for Biobased Materials (AMIBM), Maastricht University, Brightlands Chemelot Campus, Urmonderbaan 22, 6167 RD Geleen, The Netherlands*

[1] *Corresponding author: yvonne.vandermeer@maastrichtuniversity.nl*

## Graphical Abstract

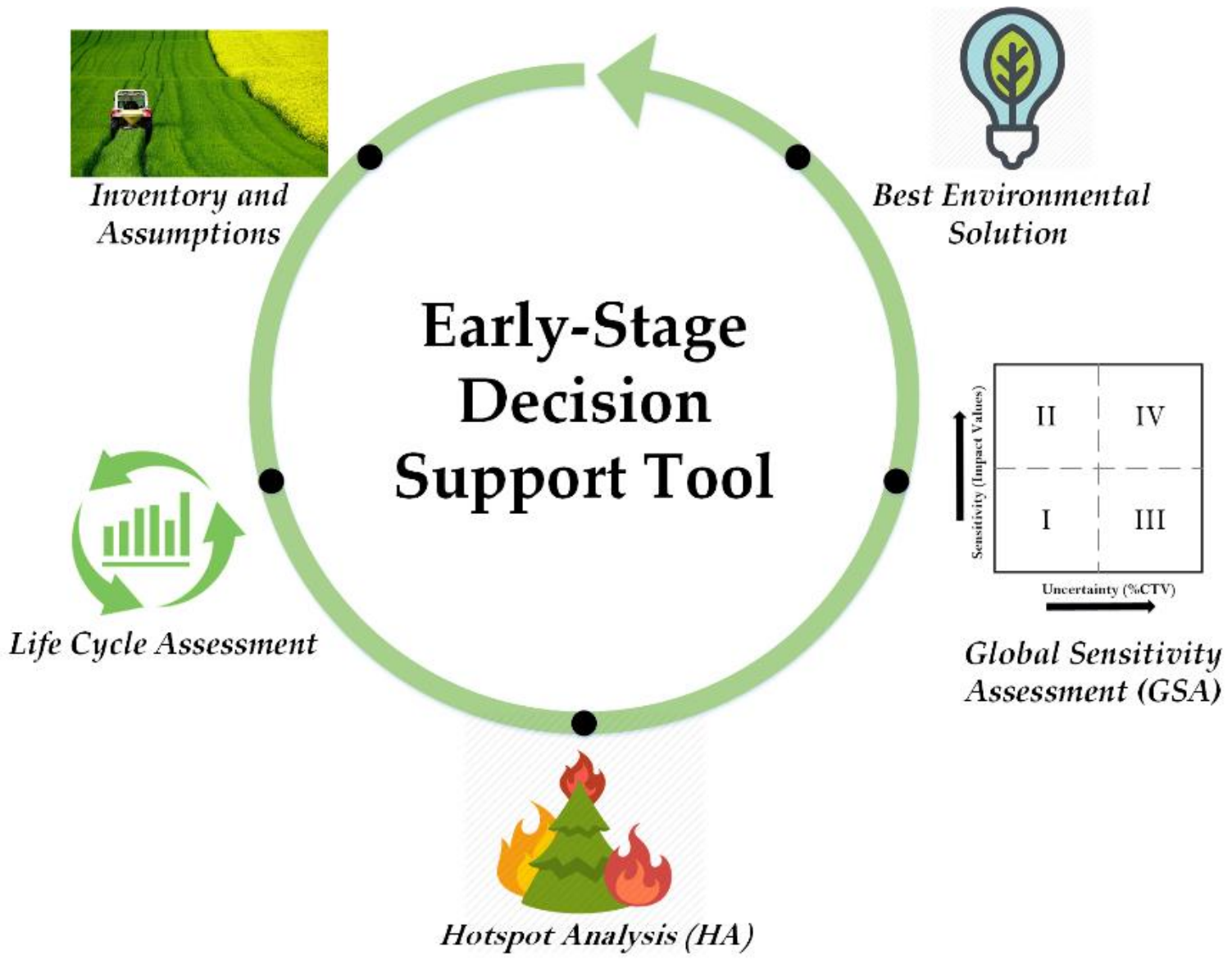

## Abstract

The transition to a low-carbon economy is one of the ambitions of the European Union for 2030. Biobased industries play an essential role in this transition. However, there has been an on-going discussion about the actual benefit of using biomass to produce biobased products, specifically the use of agricultural materials (e.g., corn and sugarcane). This paper presents the environmental impact assessment of 30% and 100% biobased PET (polyethylene terephthalate) production using EU biomass supply chains (e.g., sugar beet, wheat, and *Miscanthus*). An integral assessment between the life cycle assessment methodology and the global sensitivity assessment is presented as an early-stage support tool to propose and select supply chains that improve the environmental performance of biobased PET production. From the results, *Miscanthus* is the best option for the production of biobased PET: promoting EU local supply chains, reducing greenhouse gas (GHG) emissions (process and land-use change), and generating lower impacts in midpoint categories related to resource depletion, ecosystem quality, and human health. This tool can help improving the environmental performance of processes that could boost the shift to a low-carbon economy.

## 1. Introduction

Nowadays, synthetic polymers are one of the most functional materials due to their properties, such as durability, mechanical resistance, thermal stability, and low cost. Their relatively low price is related to the use of fossil resources (gas and crude-oil) as feedstock for their production, which also makes them suitable for mass production (Geyer et al., 2017). The worldwide supply of polymeric materials reached 359 million tons in 2018, with Europe as the second-largest producer of plastics accounting for 18.5% of global production (PlasticsEurope, 2019). The converter demand in Europe reached 51.2 million tons of polymers in 2018, from which polyethylene terephthalate (PET) covered 7.7% of the European demand, specifically in the bottle-manufacturing sector (PlasticsEurope, 2019). The main problem is the high dependency on fossil fuels to produce these plastic bottles despite the growing share of recycled PET polymer used in PET bottles (European PET Bottle Platform, 2018). However, this is a partial solution since the share of recycled PET bottles in Europe is still very low (11% - it will increase to 30% in 2030). The price of this material is not yet economically competitive with virgin PET (Arthur, 2019).

New alternatives for the current fossil feedstock for plastic production are needed. Therefore, biobased plastics have emerged as a promising alternative to mitigate the negative impact (e.g., greenhouse gas (GHG) emissions, the use of non-renewable resources) of the production of plastics using fossil resources. However, industrial biomass for biobased plastics production has opened the floor to discuss the adverse effects on eutrophication, ozone depletion, acidification, biodiversity loss, water use, and land use (Weiss et al., 2012). Bio-based plastics are made totally or partially using renewable resources instead of petroleum-based feedstock (Iwata, 2015). According to European Bioplastics (European Bioplastics, 2018), biobased plastics production was 2.11 million tons in Europe in 2019, and biobased PET accounted for 9.8% of the annual European production capacity. The high popularity of biobased PET originates from: i) the chemical and physical equivalence to the fossil counterpart, ii) it can be processed using the existing technology, and iii) the availability of bioethanol from different feedstock to produce mono ethylene glycol (MEG) as one of the main precursors (Tsiropoulos et al., 2015).

Industries are significant contributors to the adoption and expansion of new technologies. In the case of bioplastics, one of the first industries that started the transition towards using renewable feedstock in the production of plastics was the beverage industry, with the introduction of the PlantBottle™ by the Coca-Cola® company in 2009 (Anderson, 2015). This product is made of 30%wt of biobased mono ethylene glycol (MEG) and 70%wt of fossil-based terephthalic acid (TPA) (Akanuma et al., 2014). Biobased MEG can be produced through the dehydration of ethanol to ethylene, and this intermediate is used in the OMEGA process (Shell, 2018). The main advantage of this pathway is the possibility of producing ethanol from renewable resources such as sugarcane (Tsiropoulos et al., 2015). However, one of the main drawbacks is the supply chain for the production of biobased MEG. Currently, biobased MEG is produced in India from two primary sources: i) bioethanol from Indian sugarcane molasses and ii) bioethanol imported from Brazil. Next, biobased MEG is shipped to Thailand for processing to 30% biobased PET bottle-grade resin (Knutzen, 2016). The current supply chain is highly dependent on imported products from different regions of the world, and thus a regional supply to produce MEG could provide economic and environmental benefits (Knutzen, 2016). In this sense, the use of regional feedstock in Europe could help to boost the agriculture sector, mitigating adverse environmental impacts due to the optimization of supply chains, and promoting the implementation of a biobased economy. This paper explores different alternatives for the production of biobased PET using local agricultural resources in Europe as an alternative to boost local development and to decrease the dependence of imported commodities such as bioethanol. The importance of using not only first-generation feedstock (food crops grown in arable land such as sugar beet) but also second-generation feedstock (e.g., *Miscanthus*) is highlighted.

The 30% biobased PET is partly renewable since it contains 70% by weight of fossil-based TPA. Therefore, the next challenge is to produce TPA from biobased/renewable sources. Different companies, such as Danone and Nestlé Waters, joined the mission of pursuing a 100% biobased PET bottle that uses biomass feedstock that does not divert resources or land from feed to humans and animals by 2020 (Nestlé USA, 2017). However, TPA-production technologies are still under development, as explained in the paper published by (Volanti et al., 2019), where the authors performed a comparative environmental assessment of three different technologies to produce biobased TPA. This paper compares the environmental impact of the production of 30% and 100% biobased PET against the fossil-PET under different scenarios where local resources are used to produce the biobased PET. Terminologies like “bio-based” or produced using “renewable sources” do not necessarily mean that it provides environmental benefits over the fossil-based counterpart.

Life Cycle Assessment (LCA) was used to compare the different production processes based on the environmental impact of the supply chain of biobased PET production. The main goal of LCA is to quantify the amount of inputs (resources, reagents, fuels) and outputs (products, by-products, and emissions) considered to be relevant of any process, service, or product to evaluate the potential impact to the environment (European Commission and Joint Research Centre, 2010). Several authors have published studies about the environmental assessment of biobased PET using different feedstocks, system boundaries, locations, functional units. **Annex 1** presents a review of the most relevant published papers regarding the LCA of biobased PET. Most of the published papers about the production of biobased PET focus on the use of different biomass sources such as sugarcane (Semba et al., 2018; Tsiropoulos et al., 2015), corn (Akanuma et al., 2014; Shen et al., 2012), corn stover (Benavides et al., 2018) and switchgrass (Chen et al., 2016). However, they do not take into account prospective scenarios for the production of biobased PET to identify what would be the potential benefits of using local agricultural materials (e.g., sugar beet) instead of imported resources. This paper includes scenarios for a promising lignocellulosic source *Miscanthus*, that has not yet been reported. This paper focuses on the design of optimal supply chain networks, as the environmental performance of biobased PET production will be determined by the supply chain (Barbosa-Póvoa et al., 2018). There are two published papers with detailed supply chain information (Semba et al., 2018) (Tsiropoulos et al., 2015). However, the scope is limited to existing supply chains (ethanol from Brazil sugarcane and ethanol from Indian molasses) and MEG production in India (Tsiropoulos et al., 2015).

This paper aims to compare the environmental performance of the current value chain for the production of biobased PET (using sugarcane from Brazil and Indian molasses) with prospective scenarios using first- (sugar beet and wheat) and second- (*Miscanthus*) generation feedstock. Additionally, the comparison also includes the production of 100% biobased PET using sugar beet to produce TPA and determine whether renewable PET production can provide better environmental benefits than the 30% biobased and fossil-based PET. This paper proposes 12 different scenarios that include different supply chain options from the traditional network (sugarcane ethanol from Brazil to MEG production in India, and PET production in Thailand) to prospective networks (local EU supply chains). Impact categories such as global warming potential (GWP), water use (WU), aquatic eutrophication (AE), marine eutrophication (ME), terrestrial acidification (TA), and cancer toxicity (CT) for the comparison of the different scenarios were selected. Furthermore, processes with the highest contribution to the total environmental impact to produce biobased PET are identified, following the hotspot analysis (HA) approach (Hannouf and Assefa, 2018). Three cases (based on the HA) to evaluate the importance of these hotspots in the uncertainty of the results through the global sensitivity assessment (GSA) are proposed (Mutel et al., 2013). Finally, the main results and the importance of using this early-stage decision-support tool to promote a low-carbon economy transition are discussed.

This paper is structured as follows: the methodology (section 2) for the environmental assessment of biobased PET that involves: the selection of the feedstock (section 2.1), description of the processes used in the biobased PET production (section 2.2), scenario analysis (section 2.3) and LCA model (section 2.4). Furthermore, the main results (section 3) of the environmental assessment are presented. Furthermore, the discussion (section 4) compares the results of this paper with literature reports and discusses the limitations/advantages of this model.

## 2. Methodology

### 2.1. Feedstock selection

The use of biomass as a resource for producing different added-value intermediates (e.g., ethanol) and final products in the last 20 years has been widely studied (Ragauskas et al., 2006). On the one hand, ethanol has attracted the attention of industries as a building block for the production of biofuels and biochemicals in the last 15 years. The production of biobased PET depends on the technological maturity of the first-generation bioethanol production, which has been boosted due to the bonanza in corn production in the US and sugarcane in Brazil (Bordonal et al., 2018). However, increasing concerns due to the competition with food production (e.g., corn for biofuels) and the expansion of sugarcane to the native biome (Brazilian Amazon) have created the necessity to evaluate alternatives for producing bioethanol. Second-generation biomass (e.g., *Miscanthus*) has started playing an important role in the transition from food-based to advance biofuels. This paper presents a selection of different feedstocks from first- (sugarcane, sugar beet, wheat) to second-generation (*Miscanthus*) to evaluate the environmental performance of the production of biobased PET. **Table 1** summarizes the selected feedstocks, their geographical location, availability, and main uses. More details about selecting these raw materials are included in **Annex 2 of the Supplementary Material**.

**Table 1**. Summary of the feedstock used for the production of biobased PET

| Feedstock | Country | Production (Mt/year) | Uses |
|---|---|---|---|
| Sugarcane (SC) | Brazil (São Paulo) | 357.14[a] | Sugar and Ethanol |
| Sugarcane Molasses | India (Uttar Pradesh) | 5.61[b] | Ethanol |
| Sugar Beet (SugB) | Germany | 34.05[c] | Sugar and Ethanol |
| Wheat | France | 36.92[c] | Food and Ethanol |
| *Miscanthus* (Misc.) | Germany | 0.083[d] | Bioenergy |

[a] Brazilian Sugarcane Industry Association (UNICA) (UNICA (Brazilian Sugarcane Industry Association), 2019)
[b] Sugar Industry & Cane Development Department (Sugar Industry & Cane Development Department, 2019).
[c] FAOSTATS (Food and Agriculture Organization (FAO), 2019)
[d] Data based on the amount of *Miscanthus* used for bioenergy production in Germany (Fachagentur Nachwochsende Rohstoffe e.V (FNR), 2019).

## 2.2. Biobased PET production: Process description

The PET synthesis consists of the condensation polymerization of MEG (30% by weight) and TPA (70% by weight). This paper evaluates the production of two types of bottle-granulate polymer resins: i) 30% and ii) 100% biobased PET. Both polymer types include the use of biobased MEG, but the main difference between them is the source of TPA. The production of 30% biobased PET involves the use of fossil-based TPA, whereas the 100% biobased PET is produced from biobased TPA from sugar beet as feedstock. The different scenarios considered to produce bottle-grade polymer resin are summarized in **Figure 1**. Additional information about the intermediate processes to produce both 30% and 100% biobased PET is included in **Annex 2 of the Supplementary Material**.

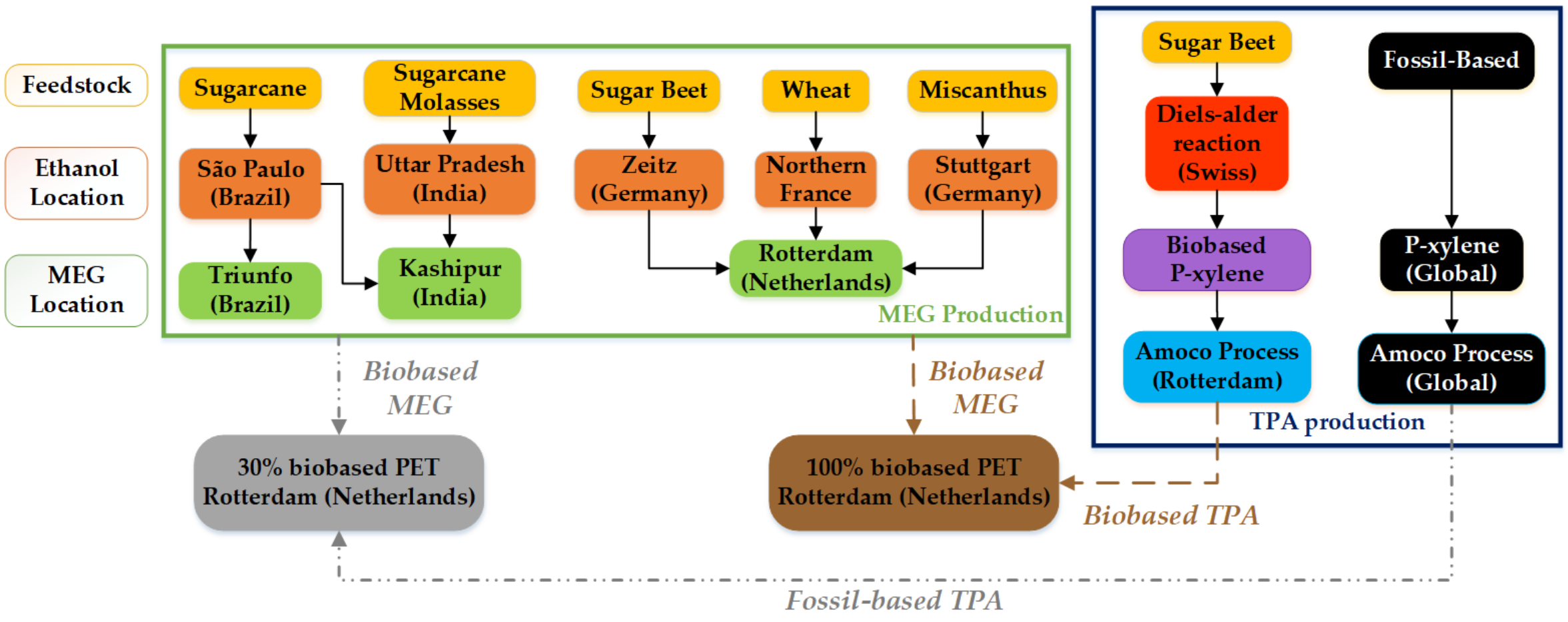

**Figure 1.** Description of the supply chain of the production of 30% and 100% biobased PET using first and second-generation biomass

## 2.3. Scenarios

A complete description of the supply chain to produce biobased PET using the selected raw materials is presented in **Table 2**. 12 scenarios using different configurations of feedstock and supply chains are proposed to evaluate the environmental performance of the biobased PET production. As mentioned in section 1, there is a lack of information regarding the biobased PET production supply chain in most of the reviewed literature publications; this provides an opportunity to propose prospective scenarios towards the deployment of biobased processes that might potentially replace the use of fossil-based materials.

The main difference between the 30% and 100% biobased PET is the location of the TPA and PET plants. For scenarios 1 to 3, the fossil-based TPA plant is located in Thailand, as previously established in the supply chain of the Coca-Cola PlantBottle (Knutzen, 2016). In the other scenarios (4 to 6), the fossil-based TPA plant was assessed as a global product with an established supply chain to trade the commodity among different locations in the world. In scenarios 7 to 12, the biobased TPA and PET are placed in the same location.

**Table 2.** Summary of the selected scenarios for the production of 30% and 100% biobased PET

| Scenario | Ethanol Feedstock | Ethanol Plant | MEG Plant | TPA Plant | Biobased-PET Plant |
|---|---|---|---|---|---|
| | | ***30% biobased PET*** | | | |
| Sc. 1[a] | Sugarcane | São Paulo (Brazil) | Kashipur (India) | Thailand[b] | Thailand |
| Sc. 2 | Sugarcane | São Paulo (Brazil)[d] | | | |
| Sc. 3 | Sugarcane Molasses | India (Uttar Pradesh) | | | |
| Sc. 4 | Sugar Beet | Zeitz (Germany) | Rotterdam (The Netherlands) | Global[b] | Rotterdam (The Netherlands) |
| Sc. 5 | Wheat | Northern region (France) | | | |
| Sc. 6 | *Miscanthus* | Stuttgart (Germany) | | | |
| | | ***100% biobased PET*** | | | |
| Sc. 7 | Sugarcane | São Paulo (Brazil) | Kashipur (India) | Rotterdam (The Netherlands)[c] | Rotterdam (The Netherlands) |
| Sc. 8 | Sugarcane | São Paulo (Brazil)[d] | | | |
| Sc. 9 | Sugarcane Molasses | India (Uttar Pradesh) | | | |
| Sc. 10 | Sugar Beet | Zeitz (Germany) | Rotterdam (The Netherlands) | | |
| Sc. 11 | Wheat | Northern region (France) | | | |
| Sc. 12 | *Miscanthus* | Stuttgart (Germany) | | | |

[a] Current supply chain to produce Coca-Cola PlantBottle (Knutzen, 2016)

[b] Location of the fossil-based TPA production

[c] Location of the biobased TPA production

[d] In this scenario, ethylene is produced in Brazil by Braskem (de Andrade Coutinho et al., 2013) and then is shipped to India for further conversion.

## 2.4. Life Cycle Assessment of Biobased PET

Life Cycle Assessment (LCA) methodology quantifies the environmental impacts attributable to the life cycle of biobased PET production. It encompasses the material and energy flows within the system boundaries and calculates relevant impacts generated by each unit process. SimaPro (Pré Consultants, the Netherlands) was used as software to perform the LCA, following the ISO 14040 and 14044 Standards (ISO, 2006). The mass and energy flows were modeled through the inventory data in the Ecoinvent database V3.4 (Ecoinvent, Switzerland). A hybrid approach involving attributional datasets from the Ecoinvent database (attributional approach), and system expansion as allocation method (consequential approach) was used. The results from the environmental assessment are used to determine the hotspots of biobased PET production. The sensitivity and uncertainty analyses are used to identify environmental improvement solutions for some of the proposed scenarios.

### *2.4.1. System Boundaries*

The assessment follows a cradle-to-gate approach, starting with the feedstock extraction, including the intermediate production facilities, and ending at the biobased PET production, including transportation to Rotterdam's port. Feedstock extraction includes agrochemical application (fertilizers, pesticides), crop management activities (plowing, planting, fertilization, harvesting, among others), and biomass transportation to the sugar mill. The intermediate facilities comprise the mass and energy balances for ethanol, MEG, TPA, and PET production along with the supply chain logistics within the different facilities. The transportation of biomass and intermediate products to the PET facility using different transportation modes such as trucks and ships was modeled, depending on the supply chain network of the scenario (more details about distances and transportation modes can be found in **Annex 3** of the Supplementary Material). A graphical description of the system boundaries considered in this study for both the production processes to obtain 30% and 100% biobased PET is presented in **Figure 2**.

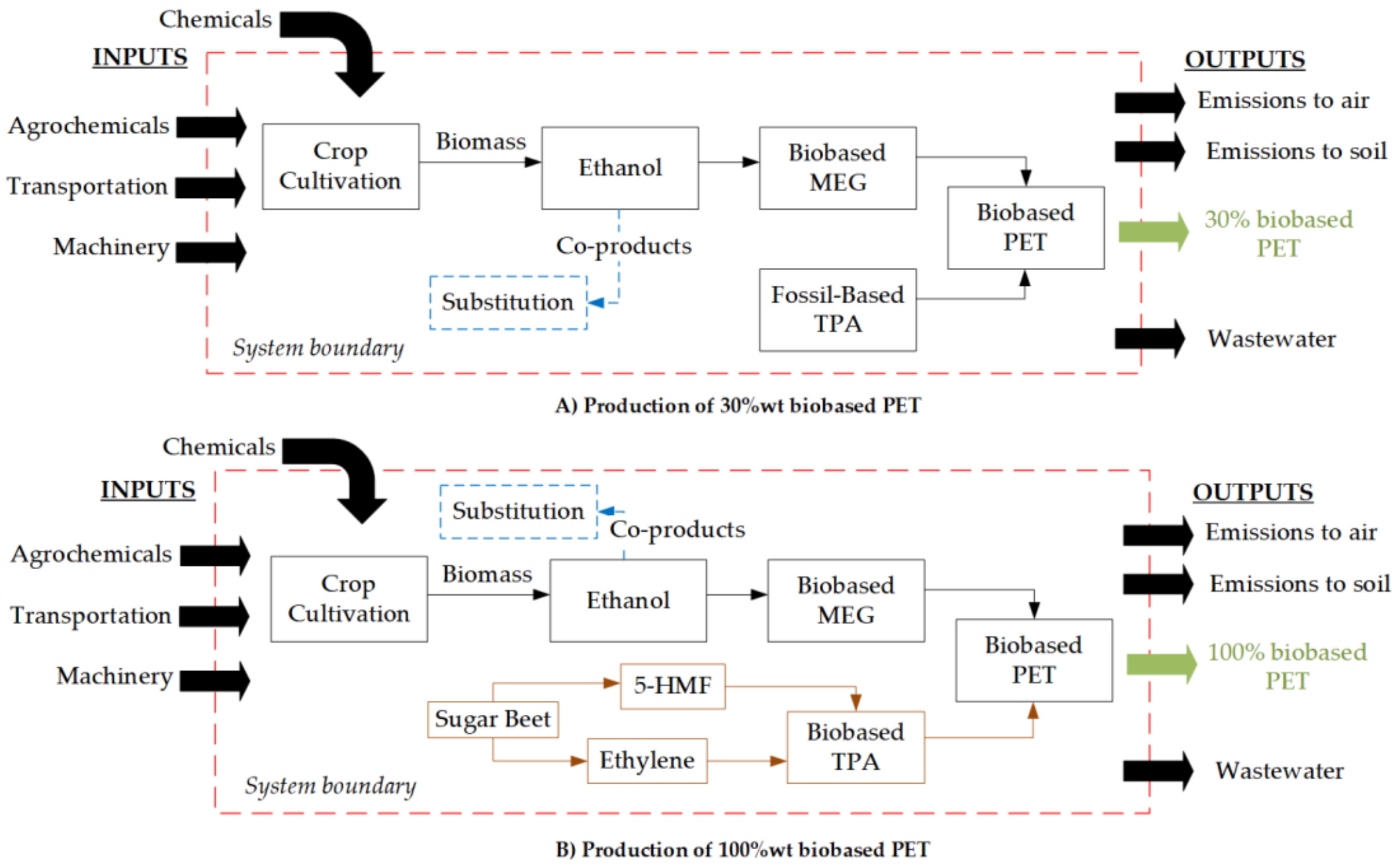

**Figure 2.** System boundaries for the LCA of the production of 30% and 100% biobased PET

### 2.4.2. *Functional Unit*

The functional unit selected in this work is *"the production of 1 kg of bottle-grade PET resin with the same chemical properties as the bottle-grade fossil PET resin delivered at the factory gate"*. This functional unit allows the comparison of the environmental impact of both fossil and biobased PET for the same application (pre-form PET bottles manufacturing).

### 2.4.3. *Allocation*

Allocation is a useful method to distribute the environmental contribution of a process or product among the main products and co-products (Cherubini and Strømman, 2011). Due to the multiple product portfolios, the production of ethanol requires the use of allocation methods. This paper uses the substitution method (system expansion) to allocate the environmental impact of the bioethanol production between the main product and co-products.

### *2.4.4. Assumptions*

Two of the critical aspects of the discussion on the environmental benefits from the production of biobased materials are Land Use Change (LUC) and biogenic carbon.

Greenhouse gas (GHG) emissions from LUC were calculated for each feedstock based on the procedure described in **Annex 3 of the Supplementary Material**. **Table 3** presents a summary of the GHG emissions of each feedstock for the production of 30% and 100% biobased PET.

**Table 3.** GHG emissions from LUC for the production of 30% and 100% biobased PET

| ***Feedstock*** | ***GHG LUC (kg $CO_{2-eq.}$/kg PET)*** | |
|---|---|---|
| | **30% Biobased** | **100% Biobased PET** |
| Sugarcane | 0.16 | 0.66 |
| Wheat | 0.7 | 1.20 |
| Sugar Beet | 0.17 | 0.67 |
| *Miscanthus* | -0.06 | 0.44 |

On the other hand, biogenic carbon was included in the GHG calculation of both 30% and 100% biobased PET, based on the percentage of carbon in the chemical structure of MEG, TPA, and PET. **Table 4** summarizes the biogenic carbon deducted from the GHG emissions of the 30% and 100% biobased PET.

**Table 4.** Biogenic carbon of 30% and 100% biobased PET

| **Material** | **Biogenic carbon (kg $CO_2$/kg PET)** |
|---|---|
| 30% biobased PET | 0.454 |
| 100% biobased PET | 2.298 |

### *2.4.5. Life Cycle Inventory (LCI)*

Data from different sources – available databases (e.g., Ecoinvent V3.4), literature reports, internet pages, and communication reports were collected. Primary data of the different scenarios were used when available, and secondary data were included to complete the inventory. **Annex 3 of the Supplementary Material** presents the life cycle inventory used to model all the proposed scenarios. **Figure 3** describes the modeling routine and data sources of the different systems in the LCA of biobased PET production. The data of the fossil-based processes (for comparison purposes) were taken from the Ecoinvent database (Hischier, 2007).

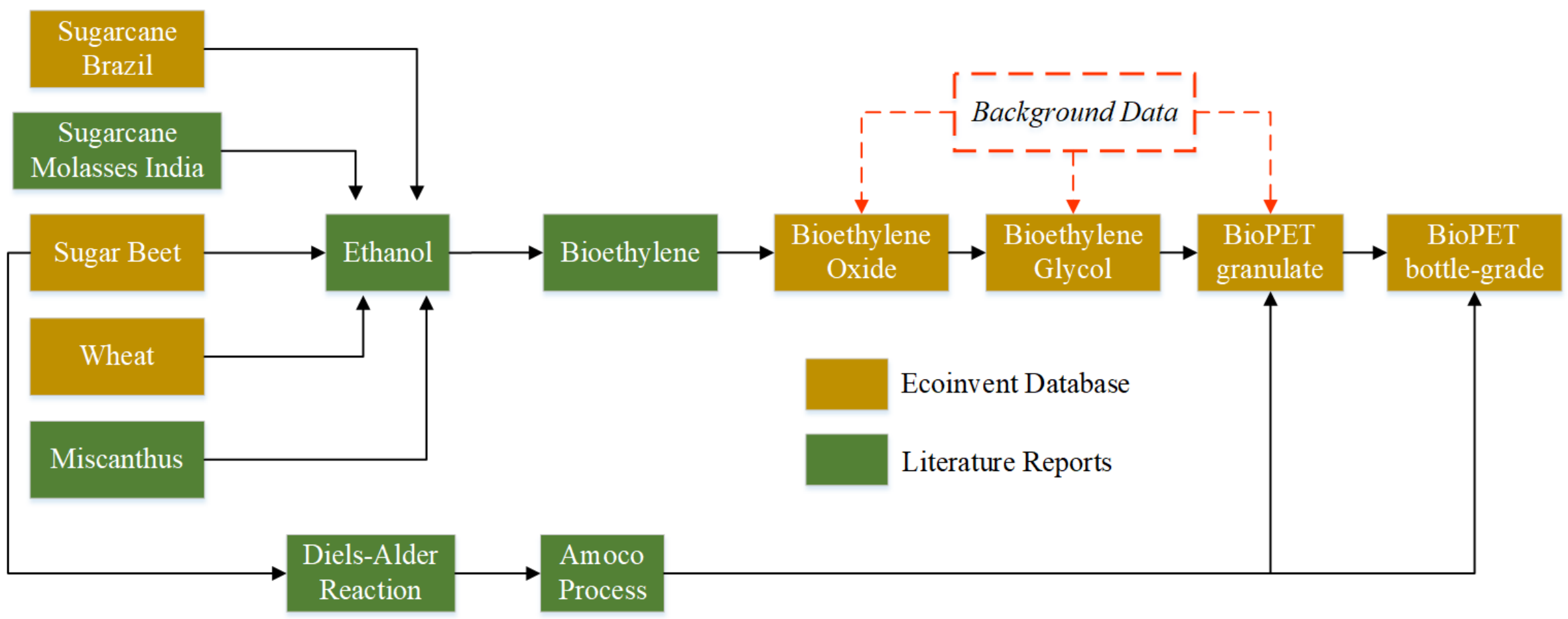

**Figure 3.** Graphical representation of the data sources for the production of biobased PET

### *2.4.6. Impact Assessment Method*

An impact assessment method that addresses the proposed scenarios from a global perspective was used to perform the LCA. IMPACT World+ method using midpoint impact categories was selected to compare the environmental impact between the proposed scenarios (Bulle et al., 2019). Midpoint categories are subject to overall fewer uncertainties than endpoint categories. Thus, the most included impact categories among LCA studies of biobased products, such as global warming potential (GWP), aquatic eutrophication (AE), marine eutrophication (ME), terrestrial acidification (TA), and cancer toxicity (CT) were used (Pawelzik et al., 2013). The water use (WU) was calculated based on the AWARE model's inventory data and characterization factors (Boulay et al., 2018).

### *2.4.7. Global Sensitivity Assessment (GSA)*

Most LCA studies address sensitivity and uncertainty separately, focusing on calculating uncertainty as an absolute value and the sensitivity as scenario analysis. This paper integrates both concepts, sensitivity and uncertainty, in a "global sensitivity assessment (GSA)" as a tool to select the best/possible environmental solutions (Hauschild et al., 2017). GSA can help understand whether the knowledge of the input data and uncertainties is sufficient to allow the decision for the supply chain to be made (Kioutsioukis et al., 2004). Three cases were selected from the hotspot analysis to implement the GSA: i) heating-source energy matrices for the production of biobased TPA, ii) fertilization method in the sugar beet cultivation iii) allocation method for the sugarcane molasses in India. A detailed description of each case is presented in **Annex 4** of the Supplementary Material. Monte Carlo (MC) simulation was used to calculate the propagation of the error in the LCA results (Heijungs and Huijbregts, 2004). The contribution to variance (CTV) was employed as metric to quantify the uncertainty of the LCA model since it estimates what percentage of the uncertainty or variance in the LCA results is caused by the assumption changes in the input parameters (Ross and Cheah, 2019). CTV was calculated as a function of the Spearman's rank-order correlation coefficient (ROCC), as previously reported by (Mutel et al., 2013).

The graphical representation introduced by (Hauschild et al., 2017) was used to represent GSA, classifying the uncertainty and sensitivity in four quadrants, as shown in **Figure 4**. The first quadrant

(I) contains the best alternatives due to the low environmental impact and uncertainty. The second (II) and third (III) quadrants contain the alternatives that could be considered as possible solutions, but either the environmental impact should improve (II) or reduce the uncertainty (III). The selection of the alternatives between quadrants II and III as the best solutions will depend on technological improvement (reduce environmental impact) and improve the input data uncertainty. Finally, the fourth quadrant (IV) contains the alternatives that are not preferred due to the high environmental impact and high uncertainty.

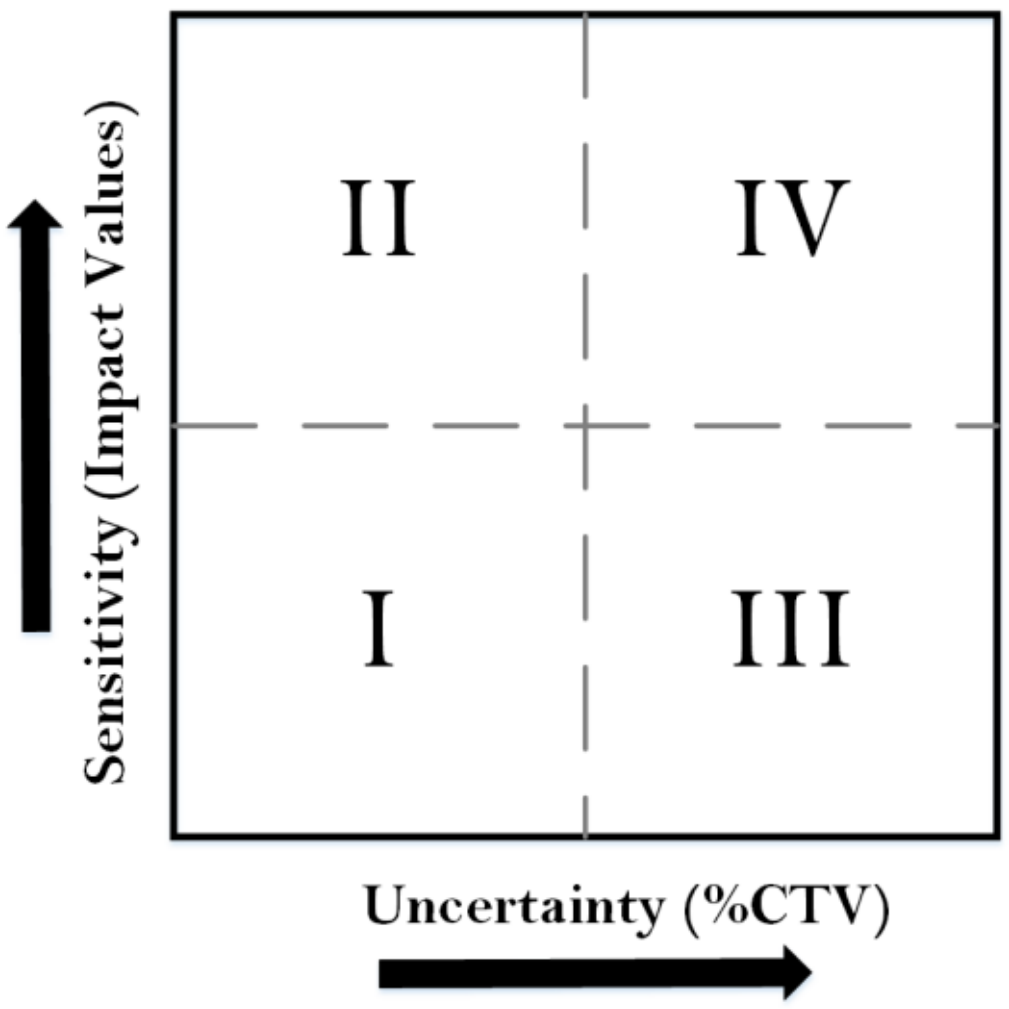

**Figure 4.** Graphical description of the global sensitivity assessment. Modified from (Hauschild et al., 2017).

## 3. Results

The results of the life cycle assessment of biobased PET production are presented in this section. GWP is used as a starting point since most of the publications reported data of this impact category. Subsequently, the results from the other selected impact categories (WU, AE, ME, TA, and CT) are summarized. For each impact category, the main contributors or hotspots of the evaluated system are determined to assess the influence of specific processes in the overall environmental impact. Finally, the uncertainty and sensitivity analysis results are introduced as a tool to select the best/possible environmental solutions for the selected study cases based on the hotspots analysis.

### *3.1. Hotspot analysis of the production of 30% and 100% biobased PET*

In general, the production of 100% biobased PET evidenced lower GHG emissions than fossil-PET (except in Sc. 11, where wheat was used as feedstock), as shown in **Figure 5**. This is possible due to the high carbon credit from biogenic storage (2.30 kg $CO_{2\text{-eq.}}$/kg PET). The production of 30% biobased PET using sugar beet (Sc. 4) and *Miscanthus* (Sc. 5) shows lower emissions for this type of material than the fossil-PET due to the carbon credit from biogenic storage and the low LUC emissions. Therefore, sugarcane used to produce MEG has a better environmental performance when sugar beet is utilized for biobased TPA production. It is a clear message to encourage developing the technology required to produce biobased TPA using local resources such as sugar beet.

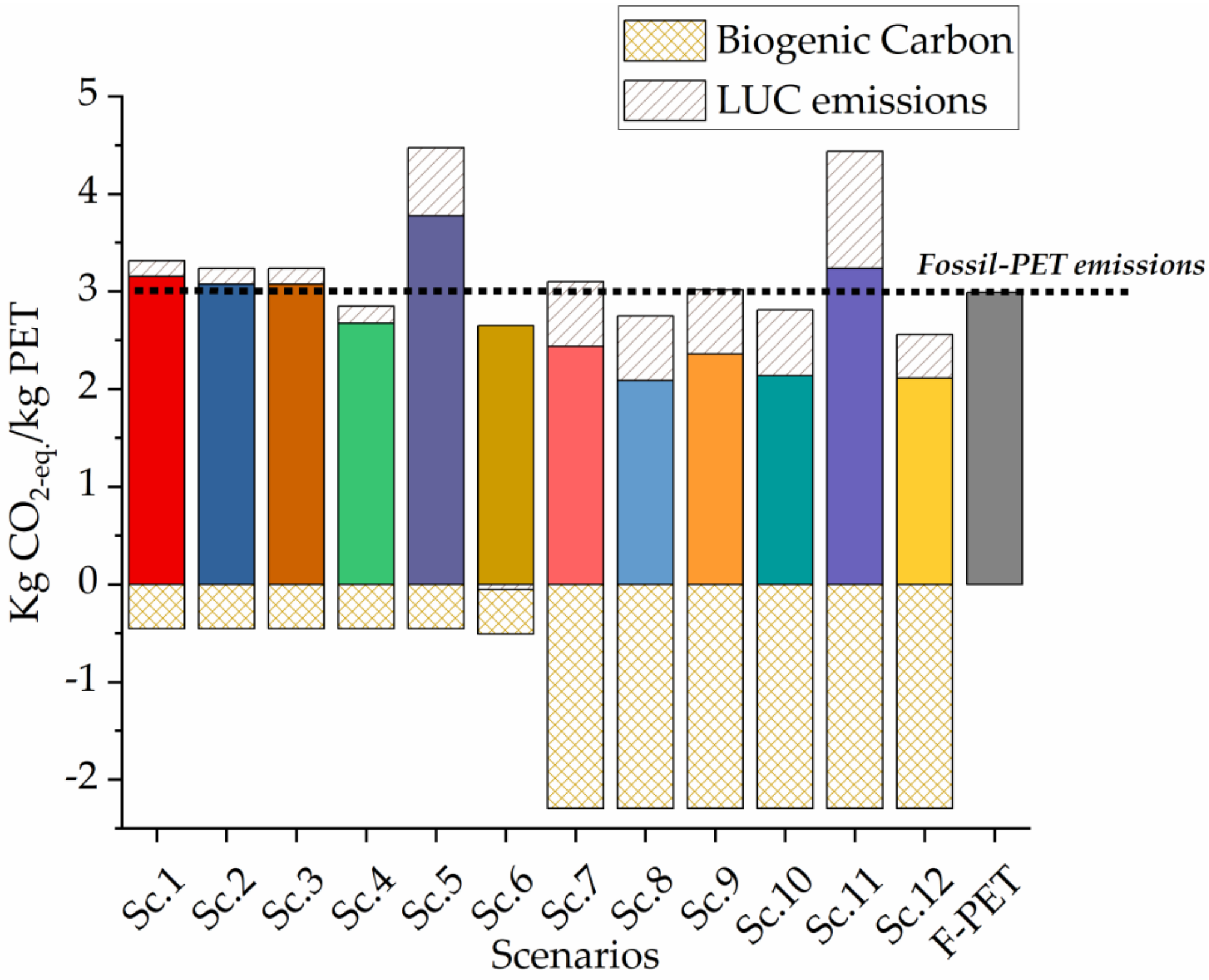

**Figure 5**. Net GHG emissions from 30% and 100% biobased PET production. *Colored bars represent the net GHG emissions*.

However, the production of 100% biobased PET using wheat (Sc. 11) has higher GHG emissions than the fossil-PET due to the high contribution of the LUC emissions. This result reassures the idea of the EC to reduce the amount of crop-based materials (e.g., maize and wheat) for the production of biochemicals. It opens the door for other types of feedstock, such as sugar beet and *Miscanthus*, to fulfill ethanol demand in the EU (Kuepker, 2018).

LUC emissions generate a debit of GHG emissions in most cases except for *Miscanthus* (Sc. 6). As explained in section 2.1, *Miscanthus* has high C-sequestration and ability to grow in degraded lands, avoiding the use of additional land for its growth and providing environmental benefits for biobased PET production. Therefore, this effect is reflected as a negative value in **Figure 5**. LUC emissions increased from 30% biobased PET to 100% biobased PET due to the additional requirement of biobased feedstock (in this case, sugar beet) for the production of p-xylene as the precursor of biobased TPA.

Despite the lower impact in the GWP category, the use of biomass for the production of biobased PET leads to higher emissions in other impact categories compared with fossil-PET, as presented in **Figures 6 – 8**. The production of 100% biobased PET performed better in terms of GWP. However, the environmental impact is higher in the AE, ME, TA, and CT impact categories than the 30% biobased PET. The use of sugarcane molasses in India to produce biobased PET (Sc.3 and Sc.9) has the highest WU due to the use of irrigation for the sugarcane, contributing up to 90% of the environmental impact, as shown in **Figure A5.4** in **Annex 5** of the Supplementary Material. WU does not change significantly for the production of 30% or 100% biobased PET due to the low WU (between 50% - 60%) in the cultivation of sugar beet as raw material for the biobased TPA production (see **Figure 6**). Regarding the TA, Sc. 4 and 10 (using sugar beet as feedstock) have the highest TA impact. Additionally, the TA impact increases when replacing fossil TPA for biobased TPA due to the use of biomass (sugar beet) for its production, which contributes to 70% of the TA impact in the TPA production (see **Figure A5.3 in the Supplementary Material**).

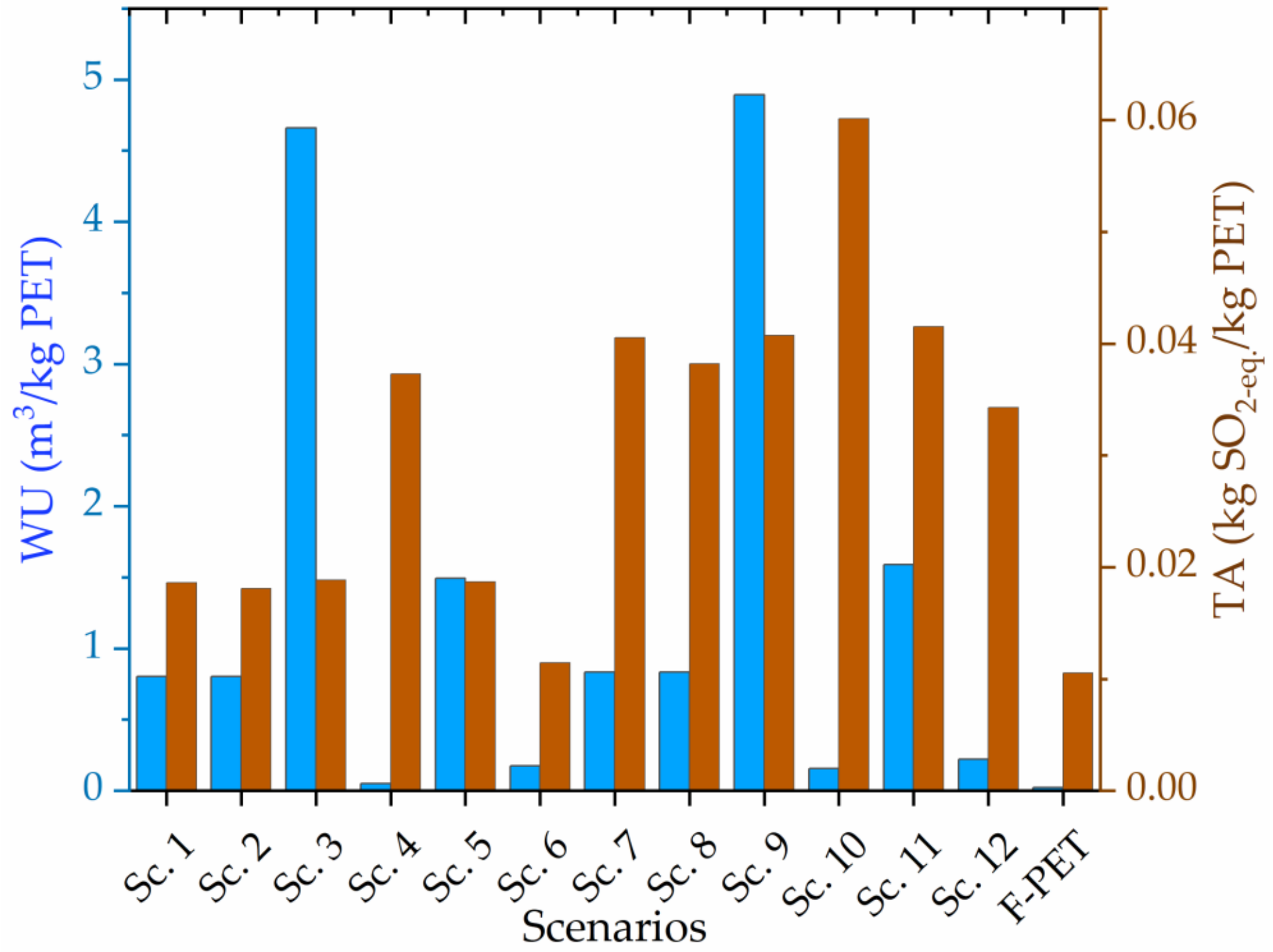

**Figure 6.** Environmental impact of the production of 30% and 100% biobased PET based on the WU and TA categories.

In the AE, the highest phosphate emissions occur in Sc. 4 and 10 due to the use of sugar beet for both biobased MEG and TPA production (see **Figure 7**). From the analysis, the hotspot in the AE impact of Sc. 4 and 10 was the use of manure for the fertilization of sugar beet (see **Figure A5.8 in Supplementary Material**). On the other hand, ME emissions from Sc. 4 and 10 were the lowest (even negative in the case of Sc. 4) due to the use of sugar beet pulp to substitute the production of mineral N-based fertilizer. The use of wheat to produce MEG showed the highest impact on the ME category (see **Figure 5.9 in Supplementary Material**).

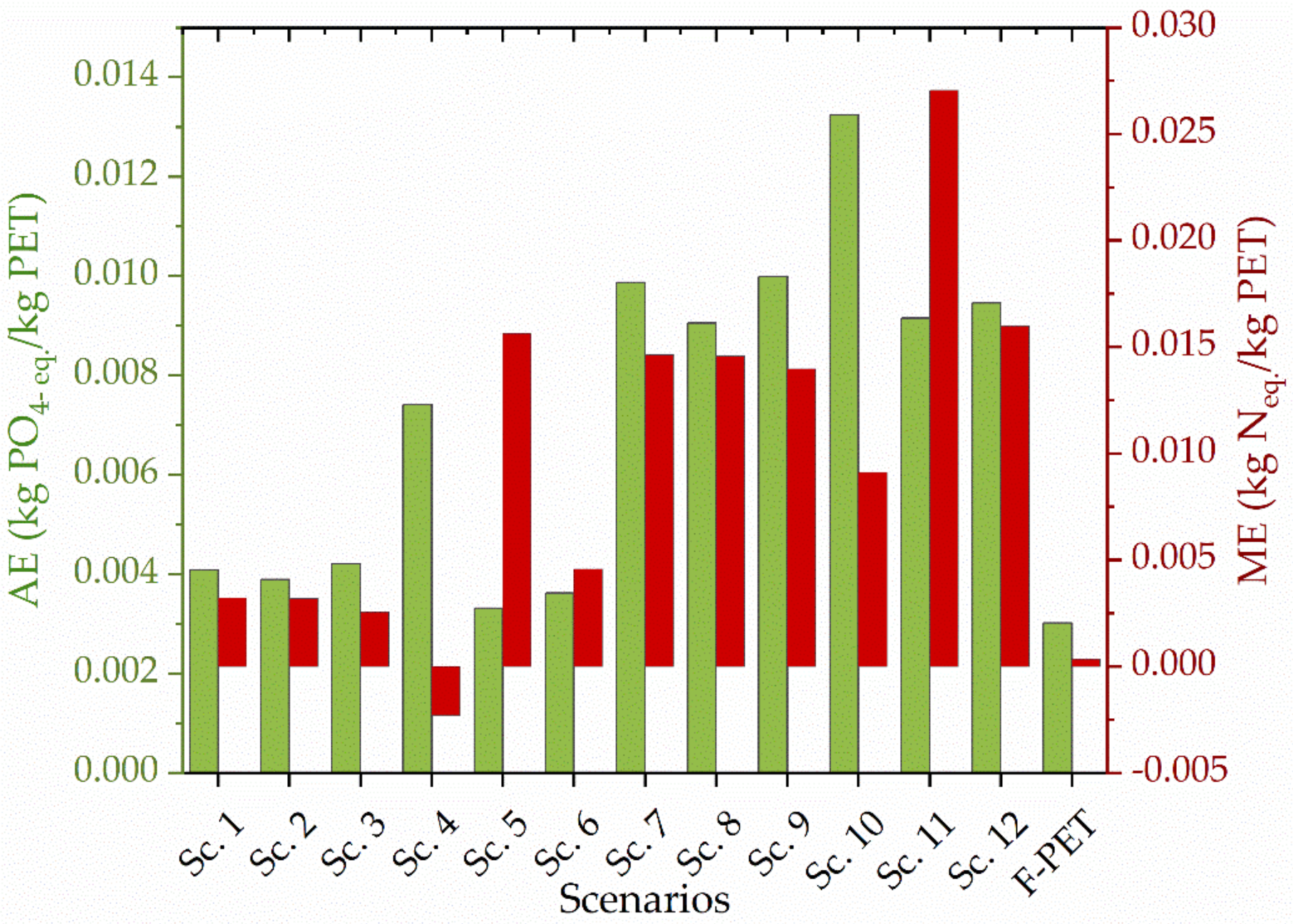

**Figure 7.** Environmental impact of the production of 30% and 100% biobased PET based on the ME and AE categories.

Following the TA and AE impact categories trend, the use of sugar beet for biobased PET production (Sc. 4 and 10) evidenced the highest CT impact (see **Figure 8**). The use of pesticides in the cultivation of sugar beet is one of the main contributors to this impact category, as explained in **Annex 2 of the Supplementary Material**).

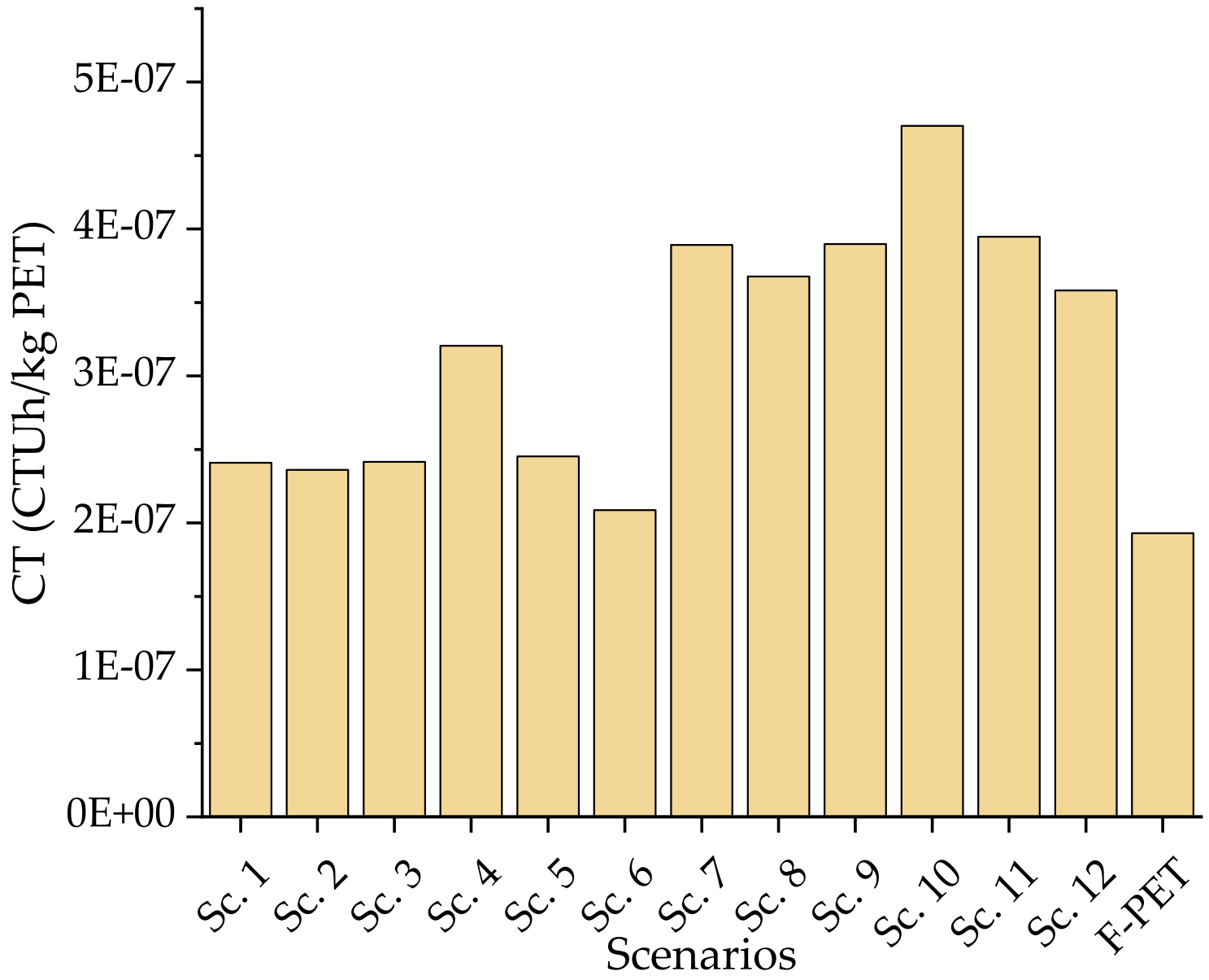

**Figure 8.** Environmental impact of the production of 30% and 100% biobased PET based on the CT category.

### *3.2. Global Sensitivity Assessment (GSA)*

**Figure 9** presents the results of the GSA of the selected heating-energy source matrices per country. Based on the results, the different heating-energy matrices are categorized in three groups: strong dependency on coal and oil, such as CZ, PL, and GR (Group 1), high dependency on natural gas, such as NL, UK, and HU (Group 2) and finally, those with strong influence from renewables such as AT, BE, FR, DE and DK (Group 3). Group 1 has the highest environmental impact in most categories except in the WU, where they also evidence the lowest CTV. On the other hand, group 2 showed a better performance in most of the impact categories; however, they provide an average-high CTV in most of the categories. Finally, group 3 seems to have a better environmental performance due to the strong input from renewables and low influence on the uncertainty of the results. Belgium (BE) seems to be a good location for biobased TPA production (from a heating-energy source perspective) due to its low environmental impact in five out of six categories, and the low CTV.

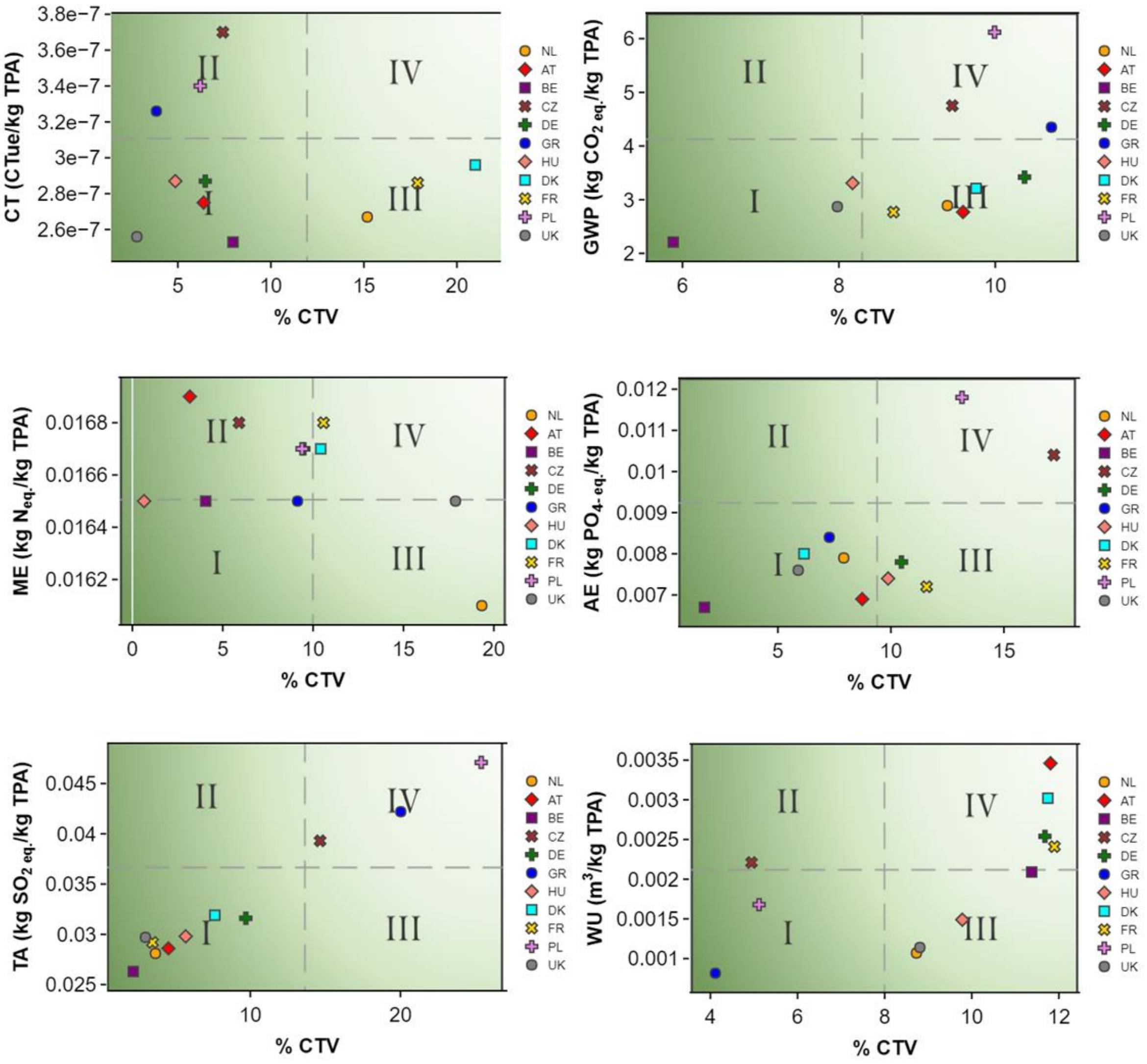

**Figure 9.** GSA of the heating-energy source in the biobased TPA production. *NL – Netherlands, AT – Austria, BE – Belgium, CZ – Czech Republic, DE – Germany, GR – Greece, HU – Hungary, DK – Denmark, FR – France, PL – Poland, UK – the United Kingdom*

The second case compared two alternatives (*Alt 1* – Use of mineral fertilizer instead of manure and *Alt 2* – Use of vinasse instead of manure) for the fertilization of sugar beet with the reference case (*Ref*) using liquid and solid manure (see, **Annex 4** in the Supplementary Material). From the GSA (**Figure 10**), *Alt 2* complies with four out of six environmental impact categories, whereas *Alt 1* only complies with one impact category (AE). The selection of the best fertilization alternative depends on the environmental objective, i.e., if the focus is on reducing GWP, *Alt 2* offers lower emissions than *Alt 1*. However, the reference case might be a better option due to the low GWP and low CTV. The *Alt 1* has low environmental performance in most of the impact categories, except in AE. *Alt 1* proposes a 50% reduction in the manure input, meaning that phosphate emissions might reduce at the expense of other impact categories (GWP, ME, TA, CT) due to additional mineral fertilizer input. Overall, the use of sugar beet vinasse as fertilizer (*Alt 2*) can improve the environmental performance of the production of sugar beet, reducing the phosphate emission (disadvantage of Sc. 5 and 100% biobased PET in **Figure 7**) and improving other impact categories. However, the uncertainty in some impact categories (CT and GWP) might require improving the input data quality.

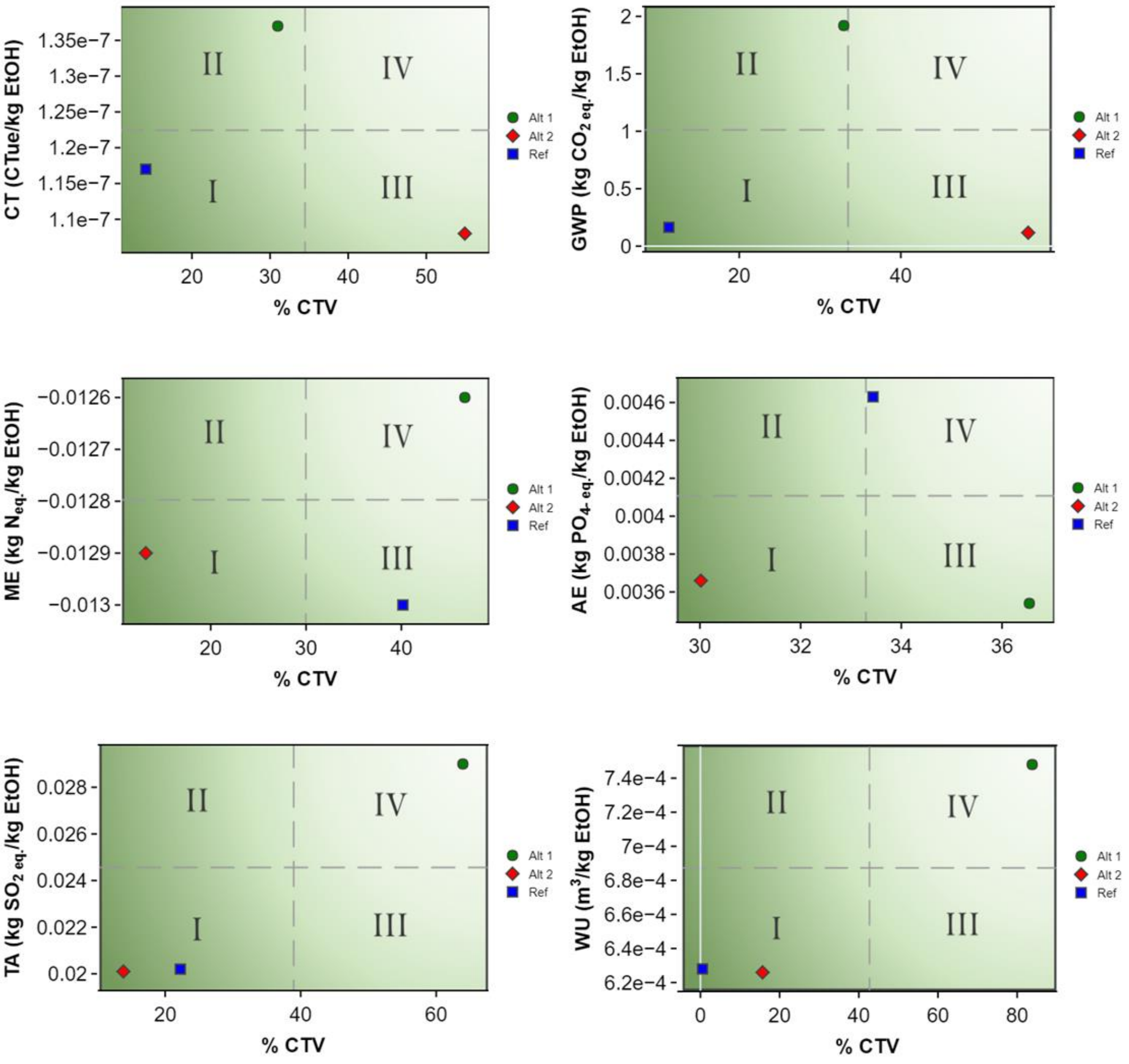

**Figure 10.** GSA of different fertilization methods in the sugar beet ethanol production.

Finally, the effect of the allocation method for the sugarcane molasses production in India as feedstock for ethanol was evaluated. The use of economic allocation gives a lower environmental impact than the mass allocation in four out of six impact categories, as presented in **Figure 11**. However, the uncertainty of the results is highly influenced by the selection of the allocation method. The use of mass allocation shows a lower CTV than the economic allocation in four out of six impact categories. The high uncertainty of the results in the economic allocation is due to the use of externalities (e.g., market prices) to distribute the environmental impact of the sugar extraction between the sugar (main product) and molasses (co-product). These results join the discussion about selecting the best allocation method that describes multiproduct processes. According to the GSA results, the mass allocation should be used (when possible), avoiding high uncertainties in the results despite the higher reported environmental impacts.

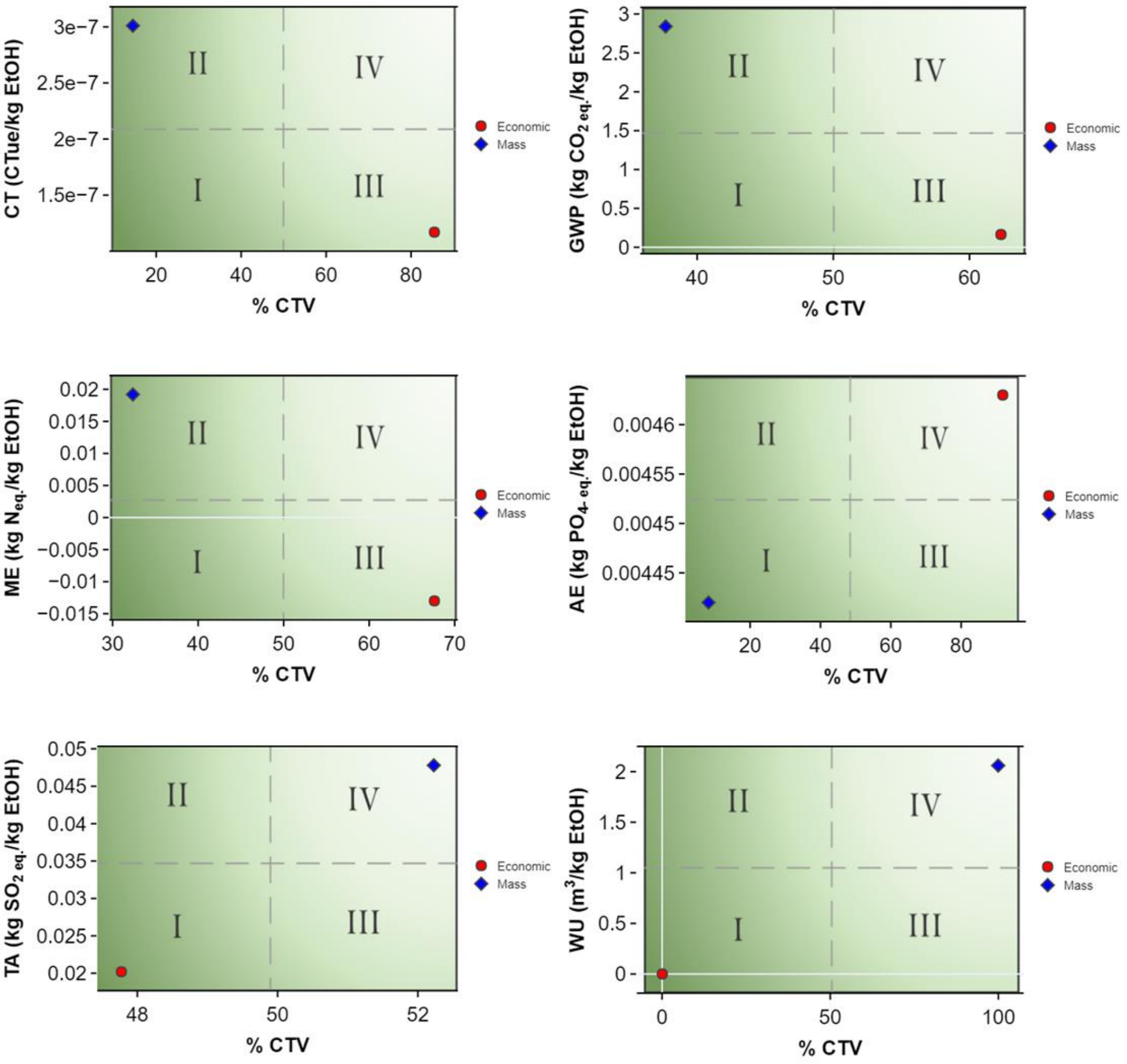

**Figure 11.** GSA of the allocation approach in the sugar molasses production in India.

## 4. Discussion

Based on the results, the implementation of supply chains that involve the use of sugar beet and *Miscanthus* for the production of 30% and 100% biobased PET should be boosted. First, sugar beet is a commonly available crop in Europe that has been affected by low sugar prices. Therefore, it is suggested its use for other purposes, such as the production of biobased chemicals. Secondly, the use of second-generation biomass (e.g., *Miscanthus* for the production of biofuels and biochemicals) is promising to mitigate the indirect LUC effects that have been linked to the use of first-generation biomass for other purposes than food. The use of sugar beet and *Miscanthus* in the production of both 30% and 100% biobased PET could mitigate GHG emissions from fossil-based PET; however, there are some consequences in other impact categories. The increasing efforts to develop the technology (process scale) to produce biobased TPA from sugar beet as one of the required monomers to produce 100% biobased PET have promising perspectives. As shown in the results, the production of 100% biobased PET can mitigate between 20% - 30% of the GHG emissions from fossil-based PET production, even using first-generation biomass (e.g., sugarcane) for the production of MEG. On the other hand, it is highly important to include indirect LUC emissions in the discussion about the use of high land intensity crops (e.g., wheat) to produce biofuels and biochemicals. Thereby, the results suggest that it is needed policies that favor the development of local biomass supply chains (e.g., Sugar beet and *Miscanthus*) to produce biofuels and biochemical alternatives to reduce the dependency on imported commodities such as sugarcane-products.

It was showed that the production of biobased PET has a lower environmental performance in comparison to fossil-PET at impact categories (midpoint level) regarding water use, human health, and ecosystem quality. Other studies have reported similar trends where the 30% biobased PET was found to have similar GWP impact than fossil-based PET while performing worse in other midpoint impact categories (particulate matter, acidification, marine eutrophication, and freshwater ecotoxicity) (European Commission, 2018). However, the reader should refer to the discussion led by (Carus et al., 2019) about the unfair comparison of fossil-based and biobased systems. There are significant differences in the technology, the economy of scale, and environmental burdens excluded from the LCA of fossil-based materials (e.g., oil spills). Whereas, all the environmental effects of the biomass use were considered in this paper. Therefore, a fair comparison between the fossil-based and biobased PET production is needed, e.g., by using prospective LCA with a commonly agreed plausible future scenario for biobased products in 2050.

### 4.1. Comparison of published LCA of the biobased PET production

**Table 5** presents the comparison of different published papers regarding GHG emissions from the LCA of biobased PET production. There are differences in the reported values from one publication to another due to the use of different feedstock, system boundaries, and model assumptions. The results of this study are similar to those from (Tsiropoulos et al., 2015) and (European Commission, 2018) for the production of 30% biobased PET. On the other hand, there are differences in the published papers and this study for the 100% biobased production. The main difference comes from the different technologies available to produce biobased TPA (isobutanol pathway and BTX pathway). However, the results in this paper are similar to those reported by (Semba et al., 2018) using the Aqueous Phase Reforming technology developed in (Virent, 2011).

**Table 5.** Comparison of different LCA for the production of biobased PET reported in the literature

| **Reference** | ***kg $CO_{2\text{-}eq.}$/kg PET*** | | **Feedstock** |
|---|---|---|---|
| | **30% Biobased PET** | **100% Biobased PET** | |
| This study | 2.63 | 2.14 | Sugar Beet |
| This study | 2.57 | 2.11 | *Miscanthus*, sugar beet |
| (Benavides et al., 2018) | 4.10 | 1.40 | Corn stover |
| (Chen et al., 2016) | ~4.20 - 6.3 | ~4.10 - 6.5 | Wood, corn, corn stover, switchgrass, wheat straw |
| (Tsiropoulos et al., 2015) | 1.93 - 2.39 | ---* | Sugarcane |
| (Semba et al., 2018) | 3.91 | 1.88 | Sugarcane, corn |
| (Akanuma et al., 2014) | ---* | 4.31 – 7.10 | Corn, wheat stover, poplar wood |
| (European Commission, 2018) | 2.19 | ---* | Sugarcane |
| (European Commission, 2018) | 2.67 | NR | Crop mix - 36% maize, 27% sugar beet and 37% wheat |

---* Values not reported.

The results of other impact categories are limited in the reviewed literature, mainly due to the different impact assessment methods used by other authors and the focus on GHG emissions as main driver for selecting biobased over fossil-based products. The report from the EC (European Commission, 2018) presents detailed results of different impact categories for the production of 30% biobased PET. However, the model choices (e.g., impact assessment method) do not allow making a direct comparison with the results of this study. Additionally, the EC did not evaluate the production of 100% biobased, and thus the results in this paper were compared with other publications. Table 6 summarizes the comparison of the results in this paper with reported data for different impact categories. There are differences between the results in this paper and the reviewed papers in most impact categories, mainly due to the selection of different feedstock and model assumptions (e.g., impact assessment method, system boundaries, and allocation methods). Despite the differences in values, the message is clear: the production of biobased PET (mostly 100% biobased PET) can reduce the GHG emissions in comparison to the fossil-PET. Nonetheless, solutions for the consequences of intensifying biomass production should be provided in terms of depletion of resources (land and water), increased use of fertilizers (especially mineral fertilizers), and human health impacts (due to the intensive use of pesticides to control plagues).

**Table 6**. Comparison of other impact categories results from different LCA studies for the production of 1 kg of biobased PET (FU)

| **Reference** | **30% Biobased-PET** | | | | **100% Biobased-PET** | | | |
|---|---|---|---|---|---|---|---|---|
| | **WU** ($m^3$/ FU) | **AE** (kg $PO_{4-eq.}$/FU) | **TA** (kg $SO_{2\ eq.}$/FU) | **CT** (CTUh/FU) | **WU** ($m^3$/ FU) | **AE** (kg $PO_{4-eq.}$/FU) | **TA** (kg $SO_{2\ eq.}$/FU) | **CT** (CTUh/FU) |
| This study (Sugar beet) | 0.052 | 0.007 | 0.0373 | 3.21E-07 | 0.157 | 0.013 | 0.0601 | 4.70E-07 |
| This study (*Miscanthus*) | 0.176 | 0.004 | 0.0114 | 2.09E-07 | 0.221 | 0.009 | 0.0343 | 3.58E-07 |
| (Benavides et al., 2018) | 0.018 | ---* | ---* | ---* | 0.041 | ---* | ---* | ---* |
| (Tsiropoulos et al., 2015) | 0.02 - 0.12 | 0.00055 - 0.00045 | ---* | 1.5E-05 – 2.0E-05[a] | ---* | ---* | ---* | ---* |
| (European Commission, 2018) | 1.1 - 3.06 | 4.9E-04 – 5.3E-06[b] | ---* | 1.97E-08 - 2.95E-07 | ---* | ---* | ---* | ---* |
| (Akanuma et al., 2014) | ---* | ---* | ---* | ---* | ---* | 0.0033 | 0.0137 | ---* |

---* Values not reported

[a] Endpoint category – Human health (DALY/kg PET)

[b] Units – kg $P_{eq.}$/kg PET

### *4.2. LCA as an early stage decision-support tool*

LCA has been widely used to determine processes and systems with the highest contribution to the environmental performance of the product or process. This assessment is often called "Hotspot Analysis (HA)" with the primary goal to prioritize actions (e.g., improvement of process configurations, reduction of waste streams) around environmental impacts (Barthel et al., 2015). This study used HA to identify the processes (e.g., biobased MEG and TPA) that contribute the most to the environmental impact of biobased PET production. Additionally, the influence of specific parameters (e.g., heating source and fertilization) and model assumptions (e.g., allocation approach) in the environmental performance of some of the "hotspots" was determined. The GSA uses the main results of the sensitivity and uncertainty analyses to select the best configurations/options having as criteria the selection of alternatives/solutions with the lowest environmental impact and CTV. The use of this combined framework – HA and GSA – can be used to propose environmental solutions at an early stage, where decision-makers should be aware of the environmental consequences of the proposed biomass-value chains while taking into account the uncertainty in the LCA results.

Three cases are introduced to show the influence of parameters and model assumptions in selecting the best alternatives/configurations. This procedure allows proposing solutions to improve the environmental performance of processes (e.g., biobased TPA, sugar beet, and sugarcane molasses) in the biobased PET production. In the first case, less polluting energy matrices (e.g., Belgium, Austria, France, Germany, Netherlands, and Denmark) should be prioritized over traditional energy sources like coal in Poland, thus promoting the phase-out of fossil-sources in the biobased TPA production. On the other hand, the selection of alternatives for the fertilization of sugar beet in Germany is key to the environmental performance of ethanol production. Finally, the selection of the allocation method provided a better understanding of the effect of model assumptions in the environmental impact of the ethanol production in India rather than provide alternatives to decision-makers. Through different cases, it was demonstrated how the GSA is a useful tool to select environmental solutions and evaluate model assumptions within production systems to improve their environmental performance. (Mutel et al., 2013) proposed a two-step GSA using first a screening method to identify the parameters with the highest contribution to the global sensitivity, and secondly the authors use the CTV to assess the relative importance of those parameters. The authors explained the importance of these GSA approaches as the inventory databases become regionalized. Along with the conclusions of (Mutel et al., 2013), GSA and HA together as an early-stage tool can be useful for decision-makers to improve the environmental performance and identify the importance of parameters and model assumptions in the uncertainty of the results. As LCA is time-consuming, this approach allows decisions at an early stage, while the identification of factors requiring more data to improve uncertainty will reduce overall analysis time. This tool was used in this paper to select the location of a production system based on the heating-energy matrix (case 1), to improve the agriculture management techniques (case 2) and the selection of the proper allocation method (case 3).

Economic purposes drive the development of biomass supply chains, and thus the economic criteria have (for now) the priority in the sustainability balance. However, the increasing concerns about the human-related impacts on the environment have promoted the introduction of externalities into economic models. These externalities (accounted as GHG emissions from processes and/or activities) can be used in an early stage decision-making process. Therefore, an integrated assessment that combines additional social and economic criteria within a framework is needed (Hannouf and Assefa, 2018). In this study, a preliminary-defined supply chain network was used. Rotterdam (the Netherlands) was established as a strategic location for biobased PET production since it is one of Europe's most important ports. However, the production costs might play an essential role in selecting the location for the biobased PET production facility (e.g., labor is inexpensive in countries like India or Thailand compared to the Netherlands). This consideration might place environmental benefits at a secondary level, prioritizing economic benefits. The next step will be developing a decision-support tool that considers environmental criteria in the design of optimal supply chains within Europe, promoting the use of local biomass (sugar beet and *Miscanthus*) for the production of biobased PET.

## 5. *Conclusions*

The key target of EU countries for 2030 is to cut 40% of greenhouse gas emissions (from 1990 levels), prioritizing the contribution of renewables in energy production. The development of strong biobased industries (e.g., bioplastics) could help mitigate the adverse impacts of fossil-sources. In this study, the production of 30% and 100% biobased PET generated lower GHG emissions (depending on the feedstock) compared to the fossil-based PET. The accounting for biogenic carbon within the system boundaries improved the environmental performance of 100% biobased PET compared to fossil-PET and even 30% biobased PET. Nonetheless, GHG emissions from land-use change might hinder the benefits of biogenic carbon, and therefore more research in this field is needed to clarify the real impact of industrial biomass cultivation in the production of biofuels and biochemicals. Moreover, other impact categories such as water use, acidification, and eutrophication showed lower environmental performance than fossil-based PET. Thereby, biobased PET is not *green*, but it *might be greener* than fossil-based PET. On the other hand, the use of local feedstock (sugar beet and *Miscanthus*) for the production of both 30% and 100% biobased PET in Europe can provide environmental benefits over conventional supply chain networks (sugarcane from Brazil or molasses in India) and other local sources with food security issues (wheat).

Global sensitivity assessment was used to compare the environmental performance and the uncertainty of different alternatives/solutions for biobased PET production, such as the heating-energy source for biobased TPA production, the fertilization of sugar beet production and the allocation method used in the production of sugarcane molasses in India. The use of hotspot analysis together with the global sensitivity assessment as an early-stage decision-making tool can help boosting the transition to a low-carbon economy by identifying and proposing environmental solutions to biobased processes/products in order to provide environmental benefits that, in the long-term, could reduce production costs. Similarly, the design of biomass supply chains for the production of biobased materials should balance the environmental and economic criteria to provide local solutions that align with the current and future policy directives.

## 6. *Acknowledgments*

This research received funding from the EU Horizon 2020 program under Marie Sklodowska-Curie Grant Agreement No. 764713, ITN Project FibreNet.